\documentclass[fleqn,10pt]{wlscirep}

\title{Charge disproportionation and nano phase separation in \textit{R}SrNiO$_4$}
\author[1]{H. Guo}
\author[1,2]{Z. W. Li }
\author[1]{C. F. Chang}
\author[1]{Z. Hu}
\author[1,3]{C.-Y. Kuo}
\author[4]{T. G. Perring}
\author[5]{W. Schmidt}
\author[6]{A. Piovano}
\author[5]{K. Schmalzl}
\author[4]{H. C. Walker}
\author[3]{H. J. Lin}
\author[3]{C. T. Chen}
\author[7,8]{S. Blanco-Canosa}
\author[9]{J. Schlappa}
\author[10]{C. Sch\"{u}{\ss}ler-Langeheine}
\author[1]{P. Hansmann}
\author[11]{D. I. Khomskii}
\author[1]{L. H. Tjeng}
\author[1,*]{A. C. Komarek }
\affil[1]{Max-Planck-Institute for Chemical Physics of Solids, N\"{o}thnitzer Str. 40, D-01187 Dresden, Germany}
\affil[2]{Institute of Applied Magnetics, Key Lab for Magnetism and Magnetic Materials of the Ministry of Education, Lanzhou University, Lanzhou 730000, People's Republic of China}
\affil[3]{National Synchrotron Radiation Research Center (NSRRC), 101 Hsin-Ann Road, Hsinchu 30076, Taiwan}
\affil[4]{ISIS Facility, STFC Rutherford Appleton Laboratory, Harwell Oxford, Didcot, OX11 0QX, United Kingdom}
\affil[5]{Forschungszentrum J\"{u}lich GmbH, J\"{u}lich Centre for Neutron Science at ILL, 71 avenue des Martyrs, 38000 Grenoble, France}
\affil[6]{Institut Laue-Langevin, Boite Postale 156X, F-38042 Grenoble Cedex 9, France} 
\affil[7]{IKERBASQUE, Basque Foundation for Science, 48013 Bilbao, Basque Country, Spain}
\affil[8]{Donostia International Physics Center, DIPC, 20018 Donostia-San Sebastian, Basque Country, Spain}
\affil[9]{European X-ray Free Electron Laser Facility GmbH, Holzkoppel 4, 22869 Schenefeld, Germany}
\affil[10]{Helmholtz-Zentrum Berlin f\"{u}r Materialien und Energie GmbH, Albert-Einstein-Str. 15, 12489 Berlin, Germany} 
\affil[11]{Physics Institute II, University of Cologne, Z\"{u}lpicher Str. 77, 50937 Cologne, Germany}
\affil[*]{Alexander.Komarek@cpfs.mpg.de}

\begin{abstract}
We have successfully grown centimeter-sized layered $R$SrNiO$_4$ single crystals under high oxygen 
pressures of 120~bar by the floating zone technique. This enabled us to perform neutron scattering 
experiments where we observe close to quarter-integer magnetic peaks below $\sim$77~K that are 
accompanied by steep upwards dispersing spin excitations. Within the high-frequency Ni-O bond stretching 
phonon dispersion, a softening at the propagation vector for a checkerboard modulation can be observed. 
Together with our spin wave simulations these observations reveal that this Ni$^{3+}$ system exhibits 
charge disproportionation with charges segregating into a checkerboard pattern within a nano phase separation scenario rather than showing a 
Jahn-Teller effect. 
\end{abstract}
\begin{document}

\flushbottom
\maketitle 
\thispagestyle{empty}

\section*{Introduction}

One of the very interesting classes of transition metal compounds showing quite unusual and rich 
properties are the rare earth (RE) perovskite nickelates \textit{R}NiO$_3$, containing Ni ions in a 
nominal low-spin 3+ state in octahedral coordination. As was shown starting with the work of Torrance 
\textit{et al.} \cite{Torrance_RNiO3},  for small RE ions there exist two transitions in these systems, from 
the high-temperature metallic orthorhombic state \textit{Pbnm} to a low-temperature insulating 
monoclinic phase \textit{P2$_1$/n}, accompanied by magnetic ordering at lower temperatures 
with a rather unusual magnetic structure with the wave vector Q=(1/4, 1/4, 1/4) (in pseudocubic notation), 
i.e. with ordering of the type $\uparrow\uparrow\downarrow\downarrow$ in all three directions. 
For NdNiO$_3$ and PrNiO$_3$ the magnetic transition and the structural transition merge into a 
common first order phase transition which simultaneously is a metal-insulator transition. And for the 
largest RE ion La, the rhombohedral LaNiO$_3$ remains metallic down to the lowest temperatures, 
and was supposed to be paramagnetic until signatures of antiferromagnetic correlations, with 
about the same magnetic propagation vector Q=(1/4, 1/4, 1/4), have recently also been discovered in 
LaNiO$_3$ single crystals grown under high oxygen pressure \cite{Guo_nc}.

It is remarkable that the Jahn-Teller distortion expected for the nominal low-spin configuration 
$t_{2g}^6$$e_g^1$ does not materialize in this \textit{R}NiO$_3$ system. This was explained
early on \cite{Mizokawa2000,Johnston_CD} by the idea of charge disproportionation of the 
type 2$\cdot$Ni$^{3+}$ $\rightarrow$ Ni$^{2+}$  + Ni$^{4+}$ which actually occurs rather on the 
ligands. The Ni$^{3+}$ and even more so the Ni$^{4+}$ in oxides have very small or negative 
charge transfer energies $\Delta$ (= E(d$^{(n+1)}$2p$^5$) - E(d$^n$2p$^6$)  )
\cite{Zaanen1985,KhomskiiB,Green2016}, so that the charge disproportionation should be  
viewed rather as 2$\cdot$(Ni$^{2+}$$\underline{L}$) $\rightarrow$  Ni$^{2+}$  +  Ni$^{2+}$$\underline{L}^2$,
where $\underline{L}$ stands for a ligand hole, which could then be termed as a valence bond 
disproportionation \cite{Bisogni2016}. This picture explains the structural transitions in 
\textit{R}NiO$_3$, at which inequivalent Ni ions appear \cite{Alonso1999}. 
Moreover, within the charge disproportionation scenario also the low-temperature magnetic structure of the $\uparrow\uparrow\downarrow\downarrow$ type \cite{Mizokawa2000}
is naturally explained. However, so far, only ‘ad-hoc’ configuration interaction models exist, 
and the phenomenon remains elusive from the point of view of \textit{ab-initio} theories \cite{subedi,seth}.
Note that the tendency to a charge, or valence bond disproportionation 
is a local property. However, there exists also an alternative 
picture \cite{BalentsA,BalentsB}, relying mainly on the Fermi-surface properties of metallic nickelates, 
which is actually a collective, not local physics.

The interesting question now arises whether the charge or valence bond disproportionation still wins 
over the Jahn-Teller distortion for Ni$^{3+}$ oxides where the local coordination is no longer $O_h$
\cite{KhomskiiA,ManganiteA}. 
For example, the layered system \textit{R}SrNiO$_4$ has a K$_2$NiF$_4$ crystal structure 
with NiO$_6$ octahedra that are elongated along the tetragonal axis. In such a case would the charge
disproportionation be absent and instead the Jahn-Teller $t_{2g}^6$$e_g^1$ be stabilized?
Electronic structure calculations in the LDA+NMTO downfolding scheme predict the 3z$^2$-r$^2$
orbital to be more stable than the  x$^2$-y$^2$ orbital by about $\sim$0.26~eV \cite{UchidaARPES,UchidaARPESb,HansmannThesis},
which is a large energy difference regarding an optical gap of $\lesssim$0.2~eV.
However, until now one could not prepare large high quality single crystals of these layered Ni$^{3+}$ 
2-1-4-nickelates. \textit{R}$_{2-x}$Sr$_x$NiO$_4$ have so far been studied only for doping levels $x$ 
away from the $x$=1 pure Ni$^{3+}$ situation. It is known, for example, that for $x \ll 1$ that 
there is Ni$^{2+}$/Ni$^{3+}$ charge order which modulates the spin structure 
\cite{Tranquada_1994,Tranquada_1994_2,Tranquada_1996}. At half-doping, i.e. for  $x$~=~0.5, 
a Ni$^{2+}$/Ni$^{3+}$ checkerboard charge ordering occurs \cite{Yoshizawa_2000} similar to the 
isostructural cobaltates \cite{Drees1,Drees2,Guo_rrl,Li_SR,Li_2016,NanoCobaltates}. This checkerboard charge order 
survives up to $\sim$ 70\% of hole-doping \cite{Ishizaka_2003} and the materials stay insulating. 
 
Some initial studies on single crystals of highly hole-doped \textit{R}$_{2-x}$Sr$_x$NiO$_4$ exist
and report an about equal population of four different kind of states around $x$~$\sim$~1 :
Ni$^{2+}$, Ni$^{3+}$ with 3z$^2$-r$^2$ orbital occupation, Ni$^{3+}$ with x$^2$-y$^2$ orbital occupation and Ni$^{4+}$ \cite{Uchida_XAS}.
Moreover, also ARPES studies on these materials exist \cite{UchidaARPES,UchidaARPESb}.

However, so far, no detailed study of the physical properties of the compounds 
with the pure Ni$^{3+}$ oxidation state have been carried out and it is unknown whether a Jahn-Teller 
orbital occupation can be stabilized by the strongly distorted Ni-oxygen environment instead of the Ni$^{2+}$/Ni$^{4+}$ or Ni$^{2+}$/Ni$^{2+}$$\underline{L}^2$ charge disproportionation. We note that the mechanism for such a charge disproportionation phenomenon is quite distinct from that for a Ni$^{2+}$/Ni$^{3+}$ charge ordering since a disproportionation in a stoichiometric system involves an energy term related to the on-site Coulomb interaction $U$ (which needs to be compensated by another interaction energy), while for ordering of charge in a non-stoichiometric or doped material the term $U$ is not operative.

\begin{figure}[!h]
\centering  
\includegraphics[width=0.45\columnwidth]{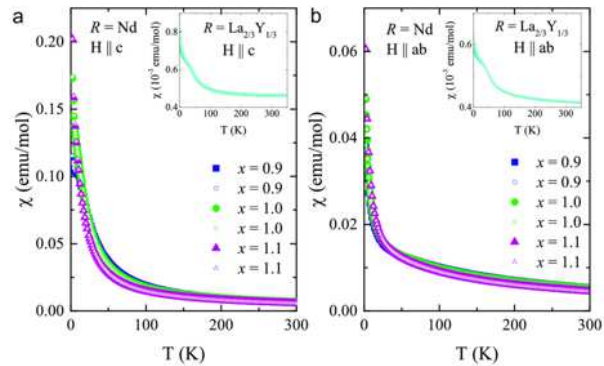}
\caption{\textbf{Magnetic susceptibility} - Temperature dependence of the magnetic susceptibility $\chi$ for Nd$_{2-x}$Sr$_x$NiO$_4$ (in the main panel) and La$_{2/3}$Y$_{1/3}$SrNiO$_4$ (in the inset). Solid/open symbols denote ZFC /FC measurements. Note, that the magnetic susceptibility of NdSrNiO$_4$ is governed by the Nd$^{3+}$ ions, which masks the signal from the Ni moments.}
\label{susceptibility}
\end{figure}

\begin{figure}[!h]
\centering  
\includegraphics[width=0.35\columnwidth]{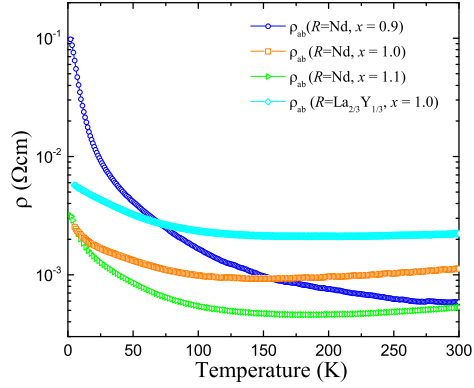}
\caption{\textbf{Electrical resistivity} - Temperature dependence of the in-plane resistivity $\rho_{ab}(T)$ for Nd$_{2-x}$Sr$_x$NiO$_4$ and La$_{2/3}$Y$_{1/3}$SrNiO$_4$.}
\label{rho}
\end{figure}
\section*{Results}

\begin{figure}[!h]
\centering 
\includegraphics[width=0.375\columnwidth]{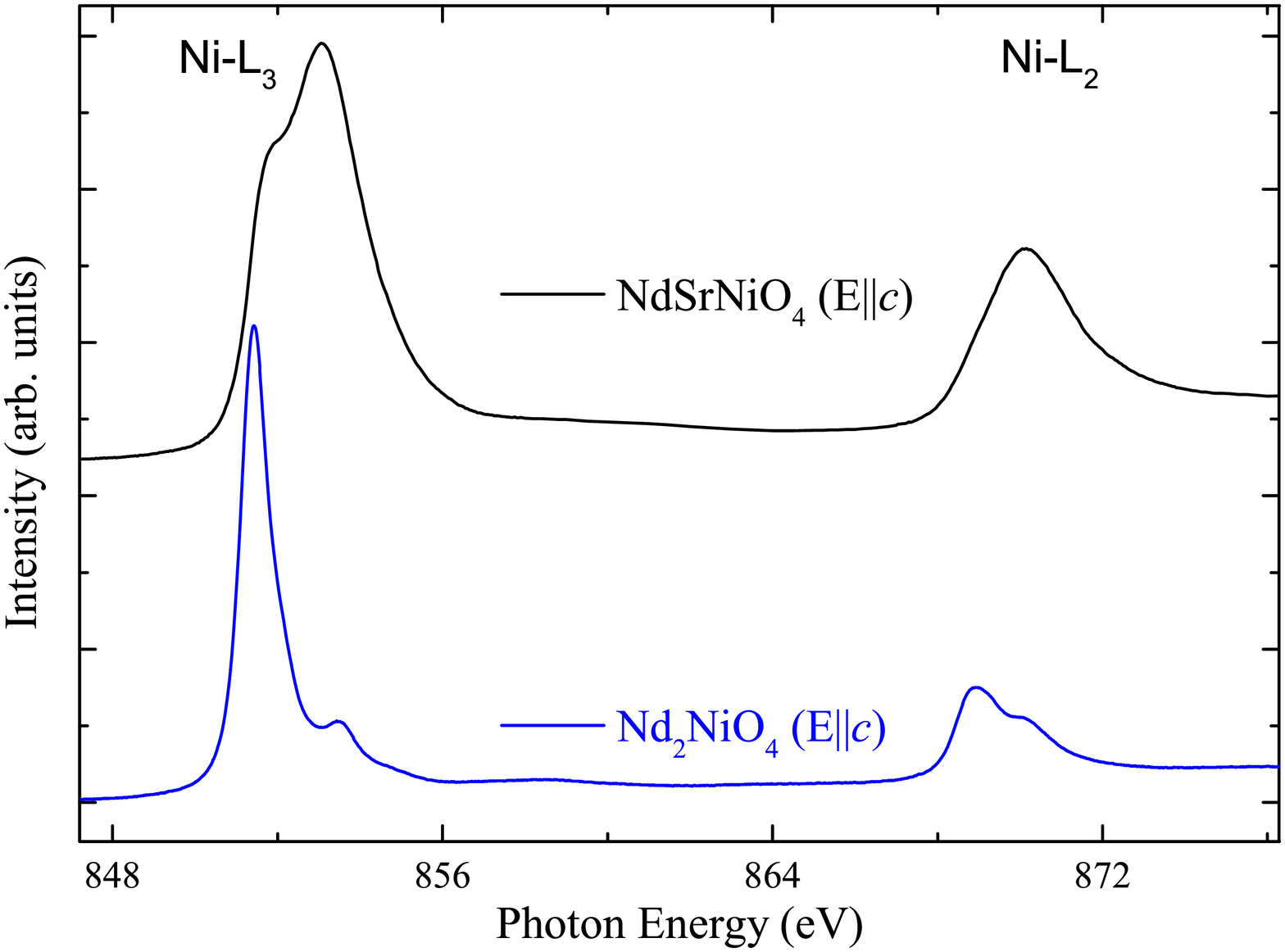}
\caption{\textbf{X-ray absorption spectroscopy } - Ni-L$_{2,3}$ XAS spectra of the NdSrNiO$_4$ 
single crystal and - for comparison - a Nd$_2$NiO$_4$ single crystal measured at 300~K. }
\label{XAS}
\end{figure}

Powder X-ray diffraction shows that our NdSrNiO$_4$ samples are impurity-free. From the refinement of the powder X-ray diffraction data we obtain a $c/a$ ratio that amounts to $\sim$3.26 and apical and basal Ni-O distances which amount to 2.058(8)~\AA\ and 1.895(1)~\AA\ respectively, thus, indicating a strongly tetragonaly distorted oxygen environment
with an apical-to-basal distance ratio of 1.09.

The magnetic susceptibility in NdSrNiO$_4$ is dominated by the Nd$^{3+}$ moments, see Fig. \ref{susceptibility}. For La$_{2/3}$Y$_{1/3}$SrNiO$_4$ with non-magnetic $R$-ions the effective moment of Ni can be determined and amounts to $\sim$0.23~$\mu_\mathrm{B}$/Ni according to Curie-Weiss fits.

The temperature dependence of the in-plane resistivity is shown in Fig.~\ref{rho}. A bad metallic behavior can be observed at high temperatures followed by a semiconducting temperature dependence at lower temperatures, thus, indicating a (gradual) metal-semiconductor transition around $\sim$150~K.

The Ni-L$_{2,3}$ x-ray absorption spectrum of our NdSrNiO$_4$ single crystal is shown in Fig.~\ref{XAS}
and is compared to that of our Nd$_2$NiO$_4$ single crystal serving as a Ni$^{2+}$ reference compound. 
It is well known that the XAS spectra at the L$_{2,3}$ edge of transition metals are highly sensitive to the 
valence state \cite{Hu2000,Burnus2008a,Burnus2008b} - an increase of the valence state of the transition 
metal ion by one causes a shift of the XAS L$_{2,3}$ spectra by one or more eV towards higher energies. 
The more than one eV higher energy shift from Nd$_2$NiO$_4$ to NdSrNiO$_4$ indicates the formal 
Ni$^{2+}$ and Ni$^{3+}$ valence states for the former and the later compound respectively. Here we would
like to note that the NdSrNiO$_4$ spectrum cannot be interpreted in terms of an ionic Ni 3d$^{7}$ 
configuration, but, rather by a coherent mixture of 3d$^{8}$ and  3d$^{8}$$\underline{L}^2$ configurations \cite{Mizokawa1995,Mizokawa2000,Johnston2014,Green2016}, where each $\underline{L}$ denotes a hole 
in the oxygen ligand. We can exclude any Ni$^{2+}$ impurities in our NdSrNiO$_4$ single crystal, otherwise 
the sharp main peak of Ni$^{2+}$ impurity spectrum would have been visible as a sharp shoulder at the 
leading edge.  

\begin{figure}[!h]
\centering 
\includegraphics[width=0.6\columnwidth]{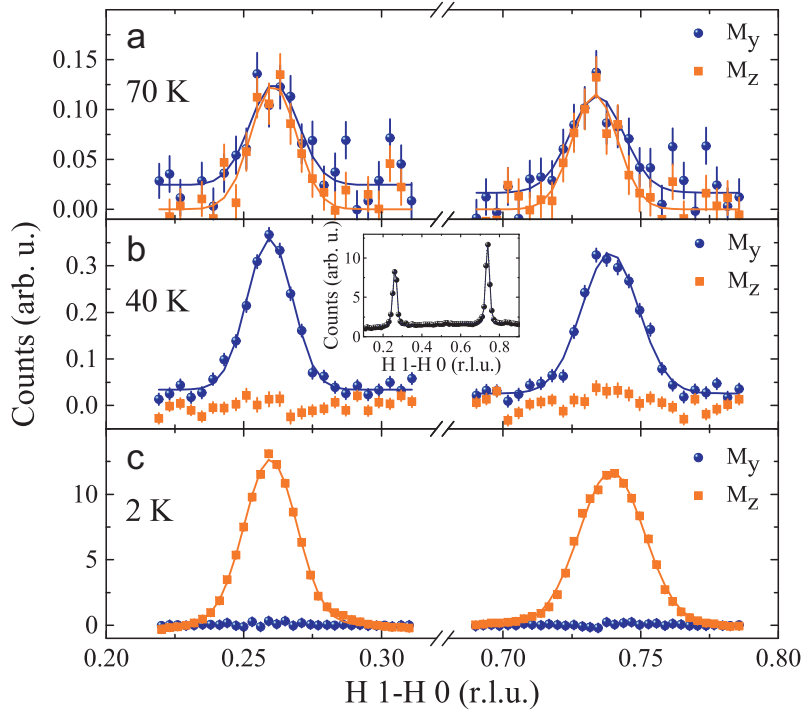}
\caption{\textbf{Neutron diffraction} - Elastic, polarized neutron scattering measurements of NdSrNiO$_4$ measured at the IN12 spectrometer
showing the extracted $M_y$ and $M_z$ components for [$H$, 1-$H$, 0] - scans at different temperatures of 
(a) 70~K, (b) 40~K and (c) 2~K. The solid curves are Gaussian fits. The inset in (b) shows 
unpolarized neutron scattering intensities measured at the IN8 spectrometer within a larger range. }
\label{elastic}
\end{figure}

\begin{figure}[!h]
\centering  
\includegraphics[width=0.43\columnwidth]{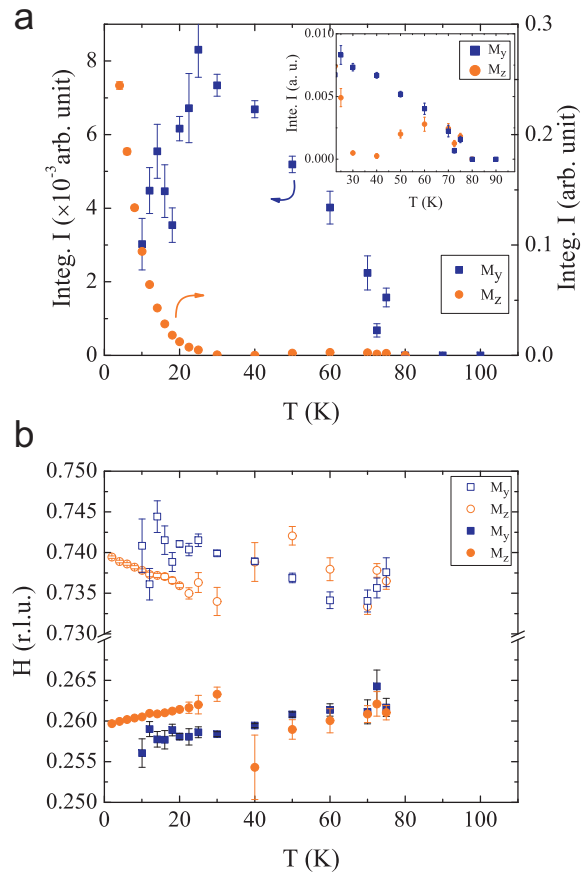}
\caption{\textbf{Analysis of polarized neutron data} - Fitting results for NdSrNiO$_4$. (a) Temperature dependence of the integrated intensities for the $M_y$ and $M_z$ components for the peak near (0.25, 0.75, 0). Note the different scales for the two components. The inset shows an enlargement of the high temperature regime. (b) Temperature dependence of the peak position at ($H$ 1-$H$ 0).}
\label{fit}
\end{figure}

\begin{figure}[!h]
\centering 
\includegraphics[width=0.43\columnwidth]{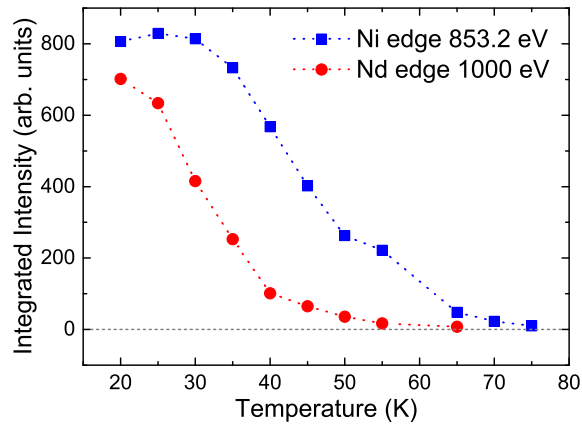}
\caption{\textbf{Resonant X-ray diffraction} - Temperature dependence of the magnetic intensities at the Nd-$M_5$ and Ni-$L_3$ resonances.}
\label{resX}
\end{figure}

\begin{figure}[!h]
\centering  
\includegraphics[width=0.5\columnwidth]{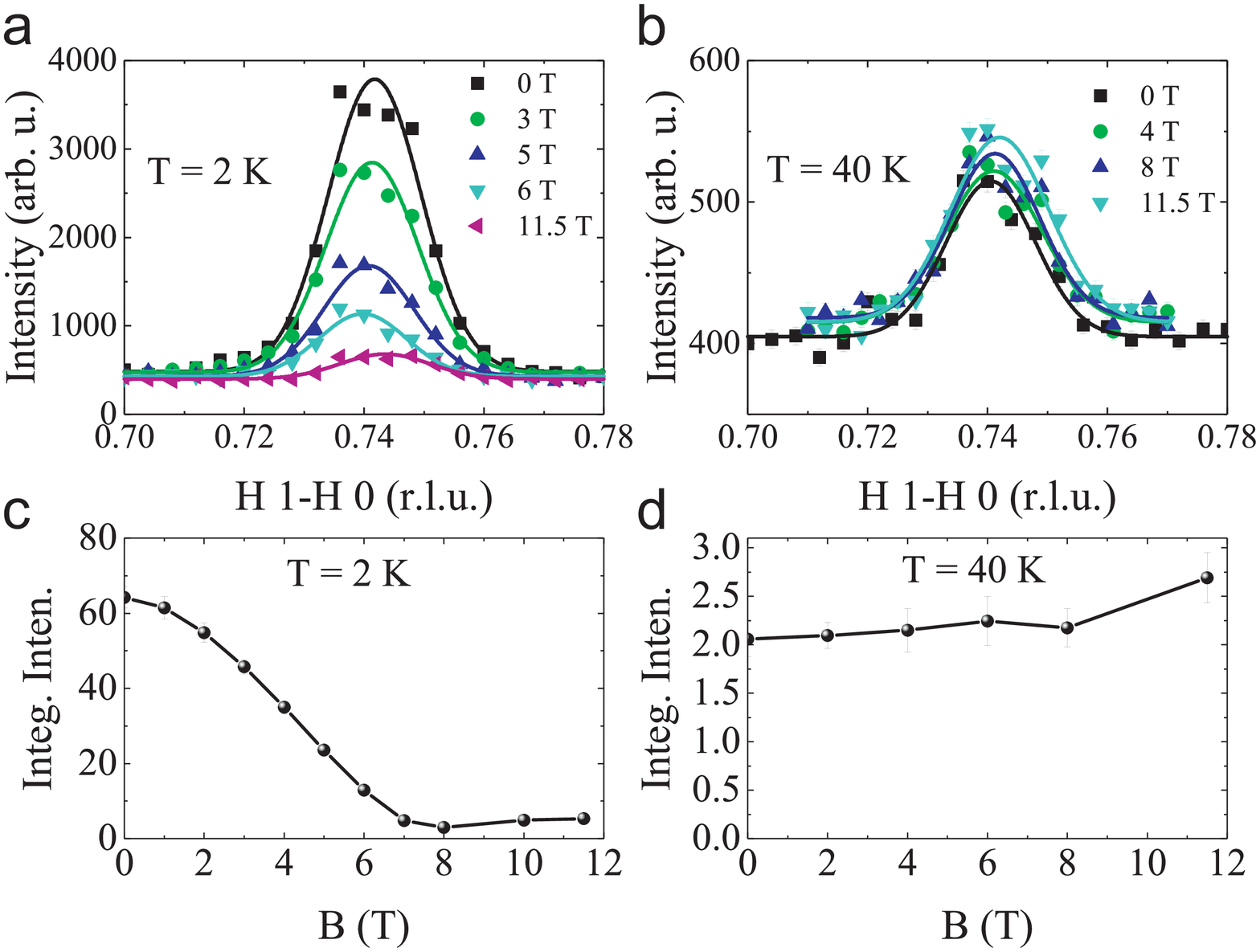}
\caption{\textbf{Magnetic field dependence} - Magnetic field dependence of the neutron intensites measured along the [$H$, 1-$H$, 0] direction at (a) 2 K and (b) 40 K in NdSrNiO$_4$. (c, d) The magnetic field dependence of the integrated intensities at 2 and 40 K, respectively.}
\label{field}
\end{figure}

Due to the availability of large single crystals we are now able to perform (polarized) neutron scattering 
experiments on \textit{R}SrNiO$_4$. As shown in Fig.~\ref{elastic}, at low temperatures (i.e. below 
T$_N$~$\sim$80~K) magnetic peaks could be detected around quarter-integer positions within the 
$H$~$K$~0 plane of reciprocal space. The magnetic origin of these peaks can be confirmed by 
polarization analysis and the extracted $M_y$ and $M_z$ components are shown  in Fig.~\ref{elastic}. Gaussian functions are fitted to the data. In NdSrNiO$_4$ the Ni moments start ordering below $\sim$77~K, see Fig.~\ref{fit}.
Close to 70~K, the intensities of the $M_y$ and $M_z$ components are comparable, thus, indicating that the magnetic moments are canted out of the $ab$ plane first. But, on further cooling below 60~K the $M_y$ component increases while the $M_z$ component almost vanishes, thus, demonstrating that the magnetic moments become confined to the $ab$ plane.
For NdSrNiO$_4$ the $M_z$ component increases substantially on further cooling below 30~K.
This can be attributed to the ordering of the Nd moments in $c$ direction at lowest temperatures. The resulting values for fitted intensities and peak positions are plotted in Fig.~\ref{fit}.

Additionally, we measured also the temperature dependence of the Nd- and Ni- contribution of the  antiferromagnetic reflections by means of resonant soft X-tray diffraction \cite{schuesslerLSNO}, see Fig.~\ref{resX}. 
These element selective measurements corroborate our findings that the Nd-moments start to order at distinctly lower temperatures than the Ni-moments. Note, that resonant X-rays are element specific and not probing the same magnetic intensity as neutrons are doing. Also short range correlations will be less integrated in this measurement due to the higher resolution which might result in a different temperature dependence compared to the neutron results. 

As shown in Fig. \ref{field}(a) and (b), the magnetic fields suppress the magnetic reflection intensities drastically at 2~K, while having almost no influence at 40~K, which can be seen more clearly from the extracted magnetic field dependence of the integrated intensities using a Gaussian fit to the data, as shown in Fig. \ref{field}(c) and (d). These results reflect the full polarization of the Nd$^{3+}$moments below the Nd-magnetic ordering temperatures in fields above 6~T whereas the (much smaller) Ni magnetic moments (that are observable at 40~K) are much less affected.

Summarizing, distinctly below T$_N$ (e.g. around 40~K) the Ni magnetic moments are 
aligned within the $ab$ planes. For NdSrNiO$_4$ also the Nd moments start 
ordering below roughly 20~K with the larger Nd moments being aligned in the $c$-direction. 
The observation of quarter integer magnetic peaks in $R$SrNiO$_4$ is compatible with a charge 
disproportionation of the nominal Ni$^{3+}$ (3d$^7$) ions into a 3d$^{8}$/3d$^{8}$$\underline{L}^2$ 
configurations. The ordering of these charges is in a checkerboard pattern.

\begin{figure}[!h]
\centering 
\includegraphics[width=0.4\columnwidth]{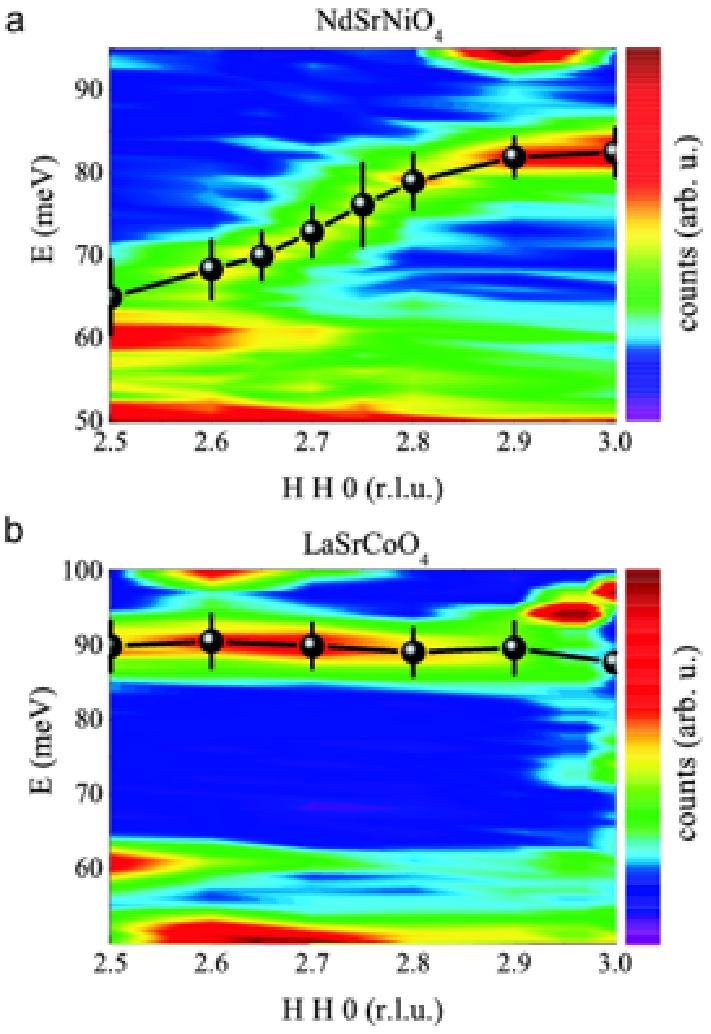}
\caption{\textbf{Bond stretching Phonons} - Phonon dispersions at 2~K measured at the IN8 spectrometer for (a)  NdSrNiO$_4$ and (b)  LaSrCoO$_4$.  }
\label{phonon}
\end{figure}

This kind of charge ordering is also in agreement with our inelastic neutron measurements of the 
longitudinal Ni-O bond stretching phonon mode, compare also Ref.~\cite{Drees1}. 
For high-T$_C$ superconducting cuprates it is known that such phonon 
softening of the Cu-O bond stretching phonon modes has been observed at the propagation vector 
of the underlying charge (stripe) order \cite{Reznik_phonon}, i.e. one would expect a bond-stretching 
phonon anomaly at the propagation vector of the underlying charge ordering propagation vector.
As can be seen in Fig.~\ref{phonon}(a), this high-frequency phonon dispersion softens for 
NdSrNi$^{3+}$O$_4$ at half-integer propagation vectors similar to that in \textit{half-doped} cobaltates 
with very robust checkerboard charge order \cite{Drees1}. In contrast to that, the same measurement 
of the \textit{pure} Co$^{3+}$ reference material LaSrCo$^{3+}$O$_4$ - see Fig.~\ref{phonon}(b) - 
reveals no such phonon softening at the zone boundary i.e. towards \textbf{Q} = (2.5~2.5~0). 
 
In NdSrNiO$_4$ the static distortions implied by this kind of charge distribution could not be detected 
in analogous elastic neutron measurements. However, these distortions might be too weak to be 
detectable possibly because of small distortions that are associated with the charge disproportionation since the charges are delocalized towards the oxygen ions. 
A similar situation might occur in LaNiO$_3$ \cite{Guo_nc}. 
Note that soft X-rays are unable to reach the required points in reciprocal space and that resonant hard x-ray scattering is unfavorable since any gain in intensity at the K-edge is marginal and will be overcompensated by the effects of fluorescence.

\begin{figure}[!h]
\centering 
\includegraphics[width=0.5\columnwidth]{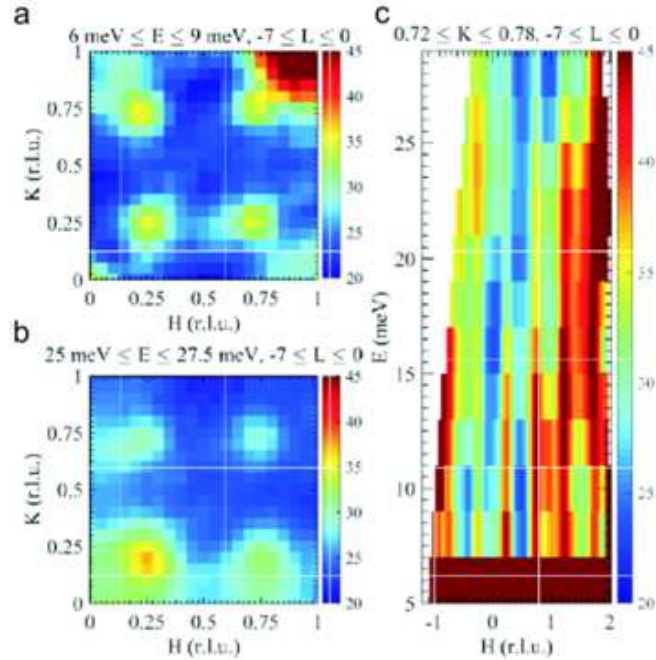}
\caption{\textbf{Spin excitations} - Magnetic excitation spectra of La$_{2/3}$Y$_{1/3}$SrNiO$_4$ measured at the  TOF spectrometer MERLIN at 6~K.}
\label{MERLIN}
\end{figure}

\begin{figure}[!h]
\centering 
\includegraphics[width=0.425\columnwidth]{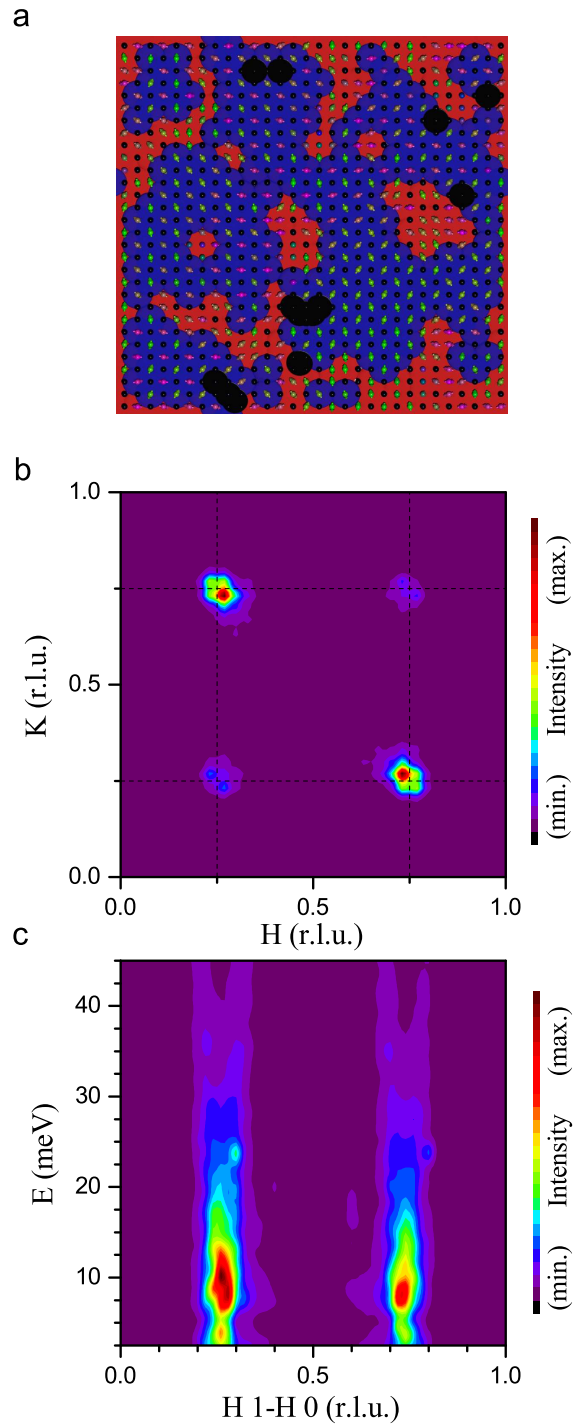}
\caption{\textbf{Nano phase separation model} - Spin wave simulations of \textit{R}SrNiO$_4$ within a nano phase separation scenario. 
(a) One of the four 30$\times$30 meshes obtained 
from MonteCarlo simulations with a similar algorithm as used for checkerboard charge ordered 
cobaltates in Ref.~\cite{Drees2}. In (b) and (c) the obtained elastic and inelastic neutron scattering 
intensities which have been calculated and averaged for four meshes are shown. Black/colored spheres: 
Ni$^{4+}$-ions/Ni$^{2+}$-ions. In this model only the Ni$^{2+}$-ions are considered since these are 
the only ions with sizable nn, nnn, etc. exchange interactions among each other. The red/blue/black 
areas indicate regions which are hosting large/small/no exchange interactions $J$/$J'$/0.
The incommensurability arises from frustration due to nano phase separation. }
\label{sim}
\end{figure}

Finally, we measured the magnetic excitation spectrum of La$_{2/3}$Y$_{1/3}$SrNiO$_4$ at  6~K (at the MERLIN 
time-of-flight spectrometer with an incident 
energy of 41~meV), see Fig.~\ref{MERLIN}. These measurements show that these nickelates are
different from the usual magnets - an upward dispersion becomes apparent which 
strongly resembles the one in highly hole-doped cobaltates \cite{Li_SR}, see  Fig.~\ref{MERLIN}(c). In the cobaltates this excitation 
spectrum could be explained by a nano phase separation model \cite{Drees2} with two relevant exchange 
interactions - one between two Co$^{2+}$-ions accross a hole ($J'$) and a much weaker one across two or 
more Co$^{3+}$-ions ($J''$ etc.). Due to the similarity with the cobaltates, one might think of a similar 
model for the nickelates, but now with larger values of the exchange interactions that will scale the entire 
spectrum to higher energies. So we assume a fully charge disproportionated nickelate system with disorder, 
where we take the nearest neighbor exchange interactions between the magnetic Ni ions to be 
$J$~$=$~$J^{100}$~$=$~-30~meV, and the exchange interaction across a non-magnetic Ni ion to be $J'$~$=$~$J^{200}$~$=$~-10~meV,
plus also a small (and less important) diagonal exchange interaction $J^{110}$~$=$~-2~meV.
Using the \textit{McPhase} program code we calculated the magnetic ground state and the corresponding 
spin wave excitations for four 30$\times$30 meshes - one of them shown in Fig.~\ref{sim}(a). The 
corresponding elastic and inelastic neutron scattering intensities of the spin correlations are shown in 
Fig.~\ref{sim}(b,c). The incommensurate magnetic peaks appear at almost the same positions as observed 
in the experiment. Furthermore, also the magnetic excitation spectrum strongly resembles the experimental 
data, compare Figs.~\ref{MERLIN} and \ref{sim}(c). Hence, a nano phase separation model for a charge 
disproportionated sample is able to explain the elastic and inelastic neutron scattering measurements in 
$R$SrNiO$_4$. Here, nano phase separation is not driven by carrier doping but by disorder within a completely 
charge disproportionated sample which inevitably leads to the creation of nanoscopic regions with large 
exchange interactions $J$ (red areas in Fig.~\ref{sim}(a)) intersected by regions with small exchange 
interactions $J'$  (blue areas in Fig.~\ref{sim}(a)) and regions with small exchange interactions that can be 
neglected (black areas in Fig.~\ref{sim}(a)). Such a scenario might be also applicable to other systems with 
checkerboard charge order with a certain degree of disorder.
Indications for nanoscale phase-separation have recently also been reported for the high-temperature superconducting cuprates  \cite{Ricci,Fratini}
which points to a possibly significant role of these effects for the physical properties of these systems.

\section*{Conclusion}
 
We succeeded in synthesizing stoichiometric Ni$^{3+}$ 2-1-4 nickelate single crystals by the high-pressure
floating zone technique. We found that this system prefers to undergo a charge disproportionation, although
it crystallizes in a tetragonal structure with strongly distorted oxygen environment of the Ni ions that would allow for a Jahn-Teller effect with an occupation 
of the $3z^2-r^2$ orbital. 
Hence, the experimental results suggest that the negative charge transfer character of the Ni$^{3+}$ ion
in these oxides provides a very strong drive for charge disproportionation to occur of the type 
2$\cdot$Ni$^{2+}$$\underline{L}$  $\rightarrow$  Ni$^{2+}$  +  Ni$^{2+}$$\underline{L}^2$, leading to the 
formation of the observed checkerboard magnetic superstructure. 
\textit{R}SrNiO$_4$ can serve as a bench mark system for further theory development to describe properly the role of holes in the oxygen band.
Finally, the magnetic correlations and the magnetic excitation spectra in these nickelates can be 
well explained by a nano phase separation model which is different from usual magnets with long range magnetic order
and points to the significance of nano phase separation for an understanding of the physical properties of these materials.

\section*{Methods}

Single crystals of NdSrNiO$_4$, La$_{2/3}$Y$_{1/3}$SrNiO$_4$ and Nd$_2$NiO$_4$ were grown by 
the floating zone technique using a high pressure mirror furnace from Scidre. Here, we studied two 
Ni$^{3+}$ systems: NdSrNiO$_4$ and La$_{2/3}$Y$_{1/3}$SrNiO$_4$. The former one because for 
\textit{R}~=~Nd there is no undesirable mixture of Ni-L$_3$ and La-M$_4$ edges in the XAS spectra,
the latter one because Nd has an undesirable large magnetic moment that might easily overshadow 
the Ni contribution. We note that the pure La system was more difficult to grow as large single crystals 
than the Y- substituted compound.  The Nd$_2$NiO$_4$ was grown to serve as a Ni$^{2+}$ reference 
system with the same crystal structure as the NdSrNiO$_4$ and La$_{2/3}$Y$_{1/3}$SrNiO$_4$. 
In order to perform floating zone growth for Nd$_{2-x}$Sr$_x$NiO$_4$ ($x$~=0.9, 1.0 and 1.1) and La$_{2/3}$Y$_{1/3}$SrNiO$_4$ starting materials of $R_2$O$_3$, SrCO$_3$ and NiO were mixed in appropriate ratios and ground thoroughly in an agate mortar followed by sintering steps at 1000$^\circ$C-1200$^\circ$C in air for several days with intermediate grindings.
The obtained composition was then pressed under $\sim$100~MPa hydrostatic pressure into a rod of about 6-8~mm in diameter and 120(20)~mm in length, which was subsequently sintered at 1000$^\circ$C in air. The floating zone growth was performed with a growth rate of 2~mm/h in a flowing O$_2$ atmosphere ($\sim$0.2~l/min) at a pressure of 100~bar, 120~ bar and 150~bar for $x$~=0.9, 1.0 and 1.1, respectively. 
 
Phase purity was confirmed by powder X-ray diffraction (XRD) measurements on ground single crystals with a 2$\theta$ step of 0.005$^o$ using Cu $K_{\alpha1}$ radiation of a laboratory X-ray source.
It was possible to describe the crystal structure properly with space group \textit{I4/mmm} for all grown samples. The refined lattice parameters and the resulting unit cell volumes are listed in table \ref{tabSup}. The linear change of the lattice parameters as a function of $x$ further confirms the successful substitution of the Nd atom by Sr atom within the Nd$_{2-x}$Sr$_x$NiO$_4$ series. 
This is also corroborated by inductively coupled plasma optical emission spectrometry (ICP-OES) measurements, see the value of $x$(ICP) in Tab.~\ref{tabSup}. Moreover, thermogravimetric measurements confirmed that the oxygen content is close to the nominal value, see Tab.~\ref{tabSup}.

\begin{table}[!h]
\centering
\caption{Growth condition and physical parameters of Nd$_{2-x}$Sr$_x$NiO$_4$ and La$_{2/3}$Y$_{1/3}$SrNiO$_4$: $v_{G}$ is the growth speed, O$_2$ pressure for the crystal growth, lattice parameters $a$, $b$, $c$ and unit cell volume from the Rietveld refinements, $\mu_{eff}^{c}$/$\mu_{eff}^{ab}$ and $\theta_{CW}^{c}$/$\theta_{CW}^{ab}$ are effective moment per formula unit and Curie-Weiss temperature determined from the Curie-Weiss fit $\chi(T) = \chi_0 + C/(T-\theta_{CW})$ to the magnetic susceptibility. $x(\mathrm{ICP})$ is measured by ICP-OES, $\delta(\mathrm{TG})$ is measured by thermogravimetric (TG) analysis.}
\label{tabSup} {
\begin{tabular}{ccccc}
\hline \hline
sample:  &        \textit{x} = 0.9     &   \textit{x} = 1.0 & \textit{x} = 1.1  &   La$_{2/3}$Y$_{1/3}$ \\
\hline
$v_{G}$\,(mm/h)     &    2    &   2   &   2    &   2   \\
O$_2$ pressure\,(bar)     &    100    &   120   &   150    &  150  \\
$a$, $b$\,({\AA})      &   3.7850(3)   &   3.7890(3)  &   3.7956(3)  & 3.7990(3)       \\
$c$\,({\AA})        &  12.352(1)   &  12.322(1)  &   12.278(1)  &  12.356(1)     \\
  volume\,({\AA}$^3$)        &  176.96(3)   &  176.90(2)  &   176.89(2)  & 178.318(3)     \\
$\mu_{eff}^{c}$\,($\mu_B$)        &   3.928(6)   &   3.777(3)  &  3.532(6)   &  0.165(1) \\
$\mu_{eff}^{ab}$\,($\mu_B$)      &   4.364(4)   &   3.942(5)  &  3.701(9)   & 0.229(1)  \\
$\theta_{CW}^{c}$\,(K)           &  -6.2(3)    &   -6.2(1)  &  -10.3(3)   &  25(1)  \\
$\theta_{CW}^{ab}$\,(K)      &    -134.3(3)    &   -105.8(4)  &  -92.3(6)   & -26(1)  \\
$x(\mathrm{ICP})$        &   0.90(2)   &  1.01(2)  &   1.12(2)   &      \\
$\delta(\mathrm{TG})$       &   0.02(2)   &   0.00(2)  &   -0.01(2)  &   \\
\hline \hline
\end{tabular}}
\end{table}

Magnetic susceptibility ($\chi$) measurements have been performed using a Quantum Design vibrating sample magnetometer (VSM)
and
the electrical resistivities have been measured with a four-probe method using the Quantum Design PPMS system.

The Ni-L$_{2,3}$ x-ray absorption spectroscopy (XAS) measurements have been performed at the 
11A beamline of the National Synchrotron Radiation Research Center (NSRRC), Taiwan. A NiO single 
crystal was measured simultaneously for energy calibration. The photon energy resolution at the Ni L$_{2,3}$ edges was set at 0.3~eV.
The spectra were recorded at 300~K using the total electron yield method.

Unpolarized elastic neutron scattering measurements were performed on the IN8 spectrometer at the 
Institut Laue Langevin (ILL) using PG monochromator and PG analyzer with fixed $k_f$~$=$~2.662\AA$^{-1}$ and two PG filters for the 
suppression of higher order contaminations. Unpolarized inelastic neutron scattering measurement 
were performed on the IN8 spectrometer at the ILL and on the MERLIN time-of-flight (TOF) spectrometer at ISIS \cite{Bewley,Russina09}. 
For the IN8 measurement, doubly focused Cu monochromator and PG analyzer were used with two 
PG filters. The MERLIN experiment has been performed in repetition-rate multiplication (RRM) mode.
Longitudinal polarized elastic neutron scattering measurements were performed on the IN12 spectrometer at the Institut Laue-Langevin (ILL) equipped with double focusing pyrolythic graphite (PG) monochromator and Heusler analyzer. The beam was polarized by a transmission polarizer in the neutron guide. 
The monochromator was set for a wave vector of 2.25~$\AA^{-1}$, and, a velocity selector was used for
suppression of higher order contamination.
The flipping ratio amounts to $\sim$22.2. For the polarization analysis, the $x$ axis is defined along the direction of \textbf{Q},
the $y$ axis is perpendicular to \textbf{Q} and within the scattering plane,
and the $z$ axis is perpendicular to the scattering plane.
Note, that we used the tetragonal setting for all our neutron measurements with lattice constant 
$a$ = $b$ $\sim$ 3.79 $\AA$ and $c$ $\sim$ 12.4 $\AA$.

Resonant soft X-ray diffraction at the Nd $3d \rightarrow 4f$ ($M_5$, 1000 eV) and Ni $2p \rightarrow 3d$ ($L_3$, 853.2 eV) resonances have been measured at the UE56/2-PGM1 beam line at BESSY II.
The data were recorded in horizontal scattering geometry with the X-rays linear polarized in the scattering plane ($\pi$-polarization). The scattered photons were detected with an in-vacuum CCD camera.

Magnetic field dependence of the antiferromagnetic correlations in NdSrNiO$_4$ was measured on the D23 diffractometer with magnetic fields applied vertically, i.e., out of the $HK$0 scattering plane.

\bibliography{NickelatesNano}

\begin{thebibliography}{10}
\expandafter\ifx\csname url\endcsname\relax
  \def\url#1{\texttt{#1}}\fi
\expandafter\ifx\csname urlprefix\endcsname\relax\def\urlprefix{URL }\fi
\providecommand{\bibinfo}[2]{#2}
\providecommand{\eprint}[2][]{\url{#2}}

\bibitem{Torrance_RNiO3}
\bibinfo{author}{Torrance, J.~B.}, \bibinfo{author}{Lacorre, P.},
  \bibinfo{author}{Nazzal, A.~I.}, \bibinfo{author}{Ansaldo, E.~J.} \&
  \bibinfo{author}{Niedermayer, C.}
\newblock \bibinfo{title}{Systematic study of insulator-metal transitions in
  perovskites r${\mathrm{nio}}_{3}$ (r=pr,nd,sm,eu) due to closing of
  charge-transfer gap}.
\newblock \emph{\bibinfo{journal}{Phys. Rev. B}} \textbf{\bibinfo{volume}{45}},
  \bibinfo{pages}{8209--8212} (\bibinfo{year}{1992}).
\newblock \urlprefix\url{https://link.aps.org/doi/10.1103/PhysRevB.45.8209}.

\bibitem{Guo_nc}
\bibinfo{author}{Guo, H.} \emph{et~al.}
\newblock \bibinfo{title}{Antiferromagnetic correlations in the metallic
  strongly correlated transition metal oxide lanio$_3$}.
\newblock \emph{\bibinfo{journal}{Nat. Commun.}} \textbf{\bibinfo{volume}{9}},
  \bibinfo{pages}{43} (\bibinfo{year}{2018}).
\newblock \urlprefix\url{https://doi.org/10.1038/s41467-017-02524-x}.

\bibitem{Mizokawa2000}
\bibinfo{author}{Mizokawa, T.}, \bibinfo{author}{Khomskii, D.~I.} \&
  \bibinfo{author}{Sawatzky, G.~A.}
\newblock \bibinfo{title}{Spin and charge ordering in self-doped mott
  insulators}.
\newblock \emph{\bibinfo{journal}{Phys. Rev. B}} \textbf{\bibinfo{volume}{61}},
  \bibinfo{pages}{11263--11266} (\bibinfo{year}{2000}).
\newblock \urlprefix\url{https://link.aps.org/doi/10.1103/PhysRevB.61.11263}.

\bibitem{Johnston_CD}
\bibinfo{author}{Johnston, S.}, \bibinfo{author}{Mukherjee, A.},
  \bibinfo{author}{Elfimov, I.}, \bibinfo{author}{Berciu, M.} \&
  \bibinfo{author}{Sawatzky, G.~A.}
\newblock \bibinfo{title}{Charge disproportionation without charge transfer in
  the rare-earth-element nickelates as a possible mechanism for the
  metal-insulator transition}.
\newblock \emph{\bibinfo{journal}{Phys. Rev. Lett.}}
  \textbf{\bibinfo{volume}{112}}, \bibinfo{pages}{106404}
  (\bibinfo{year}{2014}).
\newblock
  \urlprefix\url{https://link.aps.org/doi/10.1103/PhysRevLett.112.106404}.

\bibitem{Zaanen1985}
\bibinfo{author}{Zaanen, J.}, \bibinfo{author}{Sawatzky, G.~A.} \&
  \bibinfo{author}{Allen, J.~W.}
\newblock \bibinfo{title}{Band gaps and electronic structure of
  transition-metal compounds}.
\newblock \emph{\bibinfo{journal}{Phys. Rev. Lett.}}
  \textbf{\bibinfo{volume}{55}}, \bibinfo{pages}{418--421}
  (\bibinfo{year}{1985}).
\newblock \urlprefix\url{https://link.aps.org/doi/10.1103/PhysRevLett.55.418}.

\bibitem{KhomskiiB}
\bibinfo{author}{Khomskii, D.~I.}
\newblock \bibinfo{title}{Unusual valence, negative charge-transfer gaps and
  self-doping in transition metal compounds}.
\newblock \emph{\bibinfo{journal}{Lithuanian Journal of Physics}}
  \textbf{\bibinfo{volume}{37}}, \bibinfo{pages}{65} (\bibinfo{year}{1997}).

\bibitem{Green2016}
\bibinfo{author}{Green, R.~J.}, \bibinfo{author}{Haverkort, M.~W.} \&
  \bibinfo{author}{Sawatzky, G.~A.}
\newblock \bibinfo{title}{Bond disproportionation and dynamical charge
  fluctuations in the perovskite rare-earth nickelates}.
\newblock \emph{\bibinfo{journal}{Phys. Rev. B}} \textbf{\bibinfo{volume}{94}},
  \bibinfo{pages}{195127} (\bibinfo{year}{2016}).
\newblock \urlprefix\url{https://link.aps.org/doi/10.1103/PhysRevB.94.195127}.

\bibitem{Bisogni2016}
\bibinfo{author}{Bisogni, V.} \emph{et~al.}
\newblock \bibinfo{title}{Ground-state oxygen holes and the metal-insulator
  transition in the negative charge-transfer rare-earth nickelates}.
\newblock \emph{\bibinfo{journal}{Nat. Commun.}} \textbf{\bibinfo{volume}{7}},
  \bibinfo{pages}{13017} (\bibinfo{year}{2016}).

\bibitem{Alonso1999}
\bibinfo{author}{Alonso, J.~A.} \emph{et~al.}
\newblock \bibinfo{title}{Charge disproportionation in
  $\mathit{R}{\mathrm{nio}}_{3}$ perovskites: Simultaneous metal-insulator and
  structural transition in ${\mathrm{ynio}}_{3}$}.
\newblock \emph{\bibinfo{journal}{Phys. Rev. Lett.}}
  \textbf{\bibinfo{volume}{82}}, \bibinfo{pages}{3871--3874}
  (\bibinfo{year}{1999}).
\newblock \urlprefix\url{https://link.aps.org/doi/10.1103/PhysRevLett.82.3871}.

\bibitem{subedi}
\bibinfo{author}{Subedi, A.}, \bibinfo{author}{Peil, O.~E.} \&
  \bibinfo{author}{Georges, A.}
\newblock \bibinfo{title}{Low-energy description of the metal-insulator
  transition in the rare-earth nickelates}.
\newblock \emph{\bibinfo{journal}{Phys. Rev. B}} \textbf{\bibinfo{volume}{91}},
  \bibinfo{pages}{075128} (\bibinfo{year}{2015}).
\newblock \urlprefix\url{https://link.aps.org/doi/10.1103/PhysRevB.91.075128}.
\newblock \bibinfo{note}{PRB}.

\bibitem{seth}
\bibinfo{author}{Seth, P.} \emph{et~al.}
\newblock \bibinfo{title}{Renormalization of effective interactions in a
  negative charge transfer insulator}.
\newblock \emph{\bibinfo{journal}{Phys. Rev. B}} \textbf{\bibinfo{volume}{96}},
  \bibinfo{pages}{205139} (\bibinfo{year}{2017}).
\newblock \urlprefix\url{https://link.aps.org/doi/10.1103/PhysRevB.96.205139}.
\newblock \bibinfo{note}{PRB}.

\bibitem{BalentsA}
\bibinfo{author}{Lee, S.}, \bibinfo{author}{Chen, R.} \&
  \bibinfo{author}{Balents, L.}
\newblock \bibinfo{title}{Metal-insulator transition in a two-band model for
  the perovskite nickelates}.
\newblock \emph{\bibinfo{journal}{Phys. Rev. B}} \textbf{\bibinfo{volume}{84}},
  \bibinfo{pages}{165119} (\bibinfo{year}{2011}).
\newblock \urlprefix\url{https://link.aps.org/doi/10.1103/PhysRevB.84.165119}.

\bibitem{BalentsB}
\bibinfo{author}{Lee, S.}, \bibinfo{author}{Chen, R.} \&
  \bibinfo{author}{Balents, L.}
\newblock \bibinfo{title}{Landau theory of charge and spin ordering in the
  nickelates}.
\newblock \emph{\bibinfo{journal}{Phys. Rev. Lett.}}
  \textbf{\bibinfo{volume}{106}}, \bibinfo{pages}{016405}
  (\bibinfo{year}{2011}).
\newblock
  \urlprefix\url{https://link.aps.org/doi/10.1103/PhysRevLett.106.016405}.

\bibitem{KhomskiiA}
\bibinfo{author}{Khomskii, D.~I.}
\newblock \emph{\bibinfo{title}{Transition Metal Compounds}}
  (\bibinfo{publisher}{Cambridge Univ. Press}, \bibinfo{year}{2014}).

\bibitem{ManganiteA}
\bibinfo{author}{Wu, H.} \emph{et~al.}
\newblock \bibinfo{title}{Orbital order in la0.5sr1.5mno4: beyond a common
  local jahn-teller picture}.
\newblock \emph{\bibinfo{journal}{Phys. Rev. B}} \textbf{\bibinfo{volume}{84}},
  \bibinfo{pages}{155126} (\bibinfo{year}{2011}).

\bibitem{UchidaARPES}
\bibinfo{author}{Uchida, M.} \emph{et~al.}
\newblock \bibinfo{title}{Pseudogap of metallic layered nickelate
  ${R}_{2\ensuremath{-}x}{\mathrm{sr}}_{x}{\mathrm{nio}}_{4}$
  ($r=\mathrm{Nd},\mathrm{Eu}$) crystals measured using angle-resolved
  photoemission spectroscopy}.
\newblock \emph{\bibinfo{journal}{Phys. Rev. Lett.}}
  \textbf{\bibinfo{volume}{106}}, \bibinfo{pages}{027001}
  (\bibinfo{year}{2011}).
\newblock
  \urlprefix\url{https://link.aps.org/doi/10.1103/PhysRevLett.106.027001}.

\bibitem{UchidaARPESb}
\bibinfo{author}{Uchida, M.} \emph{et~al.}
\newblock \bibinfo{title}{Orbital characters of three-dimensional fermi
  surfaces in eu${}_{2\ensuremath{-}x}$sr${}_{x}$nio${}_{4}$ as probed by
  soft-x-ray angle-resolved photoemission spectroscopy}.
\newblock \emph{\bibinfo{journal}{Phys. Rev. B}} \textbf{\bibinfo{volume}{84}},
  \bibinfo{pages}{241109} (\bibinfo{year}{2011}).
\newblock \urlprefix\url{https://link.aps.org/doi/10.1103/PhysRevB.84.241109}.

\bibitem{HansmannThesis}
\bibinfo{author}{Hansmann, P.}
\newblock \emph{\bibinfo{title}{LDA+DMFT: from bulk to heterostructures}}
  (\bibinfo{publisher}{Technische Universität Wien}, \bibinfo{year}{2010}).

\bibitem{Tranquada_1994}
\bibinfo{author}{Tranquada, J.~M.}, \bibinfo{author}{Buttrey, D.~J.},
  \bibinfo{author}{Sachan, V.} \& \bibinfo{author}{Lorenzo, J.~E.}
\newblock \bibinfo{title}{Simultaneous ordering of holes and spins in
  la2nio4.125}.
\newblock \emph{\bibinfo{journal}{Phys. Rev. Lett.}}
  \textbf{\bibinfo{volume}{73}}, \bibinfo{pages}{1003--1006}
  (\bibinfo{year}{1994}).
\newblock \urlprefix\url{http://link.aps.org/doi/10.1103/PhysRevLett.73.1003}.

\bibitem{Tranquada_1994_2}
\bibinfo{author}{Tranquada, J.~M.} \emph{et~al.}
\newblock \bibinfo{title}{Oxygen intercalation, stage ordering, and phase
  separation in
  ${\mathrm{la}}_{2}$${\mathrm{nio}}_{4+\mathrm{\ensuremath{\delta}}}$ with
  0.05\ensuremath{\lesssim}\ensuremath{\delta}\ensuremath{\lesssim}0.11}.
\newblock \emph{\bibinfo{journal}{Phys. Rev. B}} \textbf{\bibinfo{volume}{50}},
  \bibinfo{pages}{6340--6351} (\bibinfo{year}{1994}).
\newblock \urlprefix\url{http://link.aps.org/doi/10.1103/PhysRevB.50.6340}.

\bibitem{Tranquada_1996}
\bibinfo{author}{Tranquada, J.~M.}, \bibinfo{author}{Buttrey, D.~J.} \&
  \bibinfo{author}{Sachan, V.}
\newblock \bibinfo{title}{Incommensurate stripe order in la2-xsrxnio4 with
  x=0.225}.
\newblock \emph{\bibinfo{journal}{Phys. Rev. B}} \textbf{\bibinfo{volume}{54}},
  \bibinfo{pages}{12318--12323} (\bibinfo{year}{1996}).
\newblock \urlprefix\url{http://link.aps.org/doi/10.1103/PhysRevB.54.12318}.

\bibitem{Yoshizawa_2000}
\bibinfo{author}{Yoshizawa, H.} \emph{et~al.}
\newblock \bibinfo{title}{Stripe order at low temperatures in
  ${\mathrm{la}}_{2\ensuremath{-}x}{\mathrm{sr}}_{x}{\mathrm{nio}}_{4}$ with
  $0.289\ensuremath{\lesssim}x\ensuremath{\lesssim}0.5$}.
\newblock \emph{\bibinfo{journal}{Phys. Rev. B}} \textbf{\bibinfo{volume}{61}},
  \bibinfo{pages}{R854--R857} (\bibinfo{year}{2000}).
\newblock \urlprefix\url{https://link.aps.org/doi/10.1103/PhysRevB.61.R854}.

\bibitem{Drees1}
\bibinfo{author}{Drees, Y.}, \bibinfo{author}{Lamago, D.},
  \bibinfo{author}{Piovano, A.} \& \bibinfo{author}{Komarek, A.~C.}
\newblock \bibinfo{title}{Hour-glass magnetic spectrum in a stripeless
  insulating transition metal oxide}.
\newblock \emph{\bibinfo{journal}{Nat. Commun.}} \textbf{\bibinfo{volume}{4}},
  \bibinfo{pages}{2449} (\bibinfo{year}{2013}).
\newblock \urlprefix\url{http://dx.doi.org/10.1038/ncomms3449}.

\bibitem{Drees2}
\bibinfo{author}{Drees, Y.} \emph{et~al.}
\newblock \bibinfo{title}{Hour-glass magnetic excitations induced by nanoscopic
  phase separation in cobalt oxides}.
\newblock \emph{\bibinfo{journal}{Nat. Commun.}} \textbf{\bibinfo{volume}{5}},
  \bibinfo{pages}{5731} (\bibinfo{year}{2014}).
\newblock \urlprefix\url{http://dx.doi.org/10.1038/ncomms6731}.

\bibitem{Guo_rrl}
\bibinfo{author}{Guo, H.}, \bibinfo{author}{Schmidt, W.},
  \bibinfo{author}{Tjeng, L.~H.} \& \bibinfo{author}{Komarek, A.~C.}
\newblock \bibinfo{title}{Charge correlations in cobaltates
  la$_{2-x}$sr$_x$coo$_4$}.
\newblock \emph{\bibinfo{journal}{Phys. Status Solidi RRL}}
  \textbf{\bibinfo{volume}{9}}, \bibinfo{pages}{580--582}
  (\bibinfo{year}{2015}).
\newblock \urlprefix\url{http://dx.doi.org/10.1002/pssr.201510290}.

\bibitem{Li_SR}
\bibinfo{author}{Li, Z.~W.} \emph{et~al.}
\newblock \bibinfo{title}{Incommensurate spin correlations in highly oxidized
  cobaltates la$_{2-x}$sr$_x$coo$_4$}.
\newblock \emph{\bibinfo{journal}{Sci. Rep.}} \textbf{\bibinfo{volume}{6}},
  \bibinfo{pages}{25117} (\bibinfo{year}{2016}).
\newblock \urlprefix\url{http://dx.doi.org/10.1038/srep25117}.

\bibitem{Li_2016}
\bibinfo{author}{Li, Z.~W.} \emph{et~al.}
\newblock \bibinfo{title}{Electronic and magnetic nano phase separation in
  cobaltates la$_{2-x}$sr$_x$coo$_4$}.
\newblock \emph{\bibinfo{journal}{J. Supercond. Nov. Magn.}}
  \textbf{\bibinfo{volume}{29}}, \bibinfo{pages}{727--731}
  (\bibinfo{year}{2016}).
\newblock \urlprefix\url{https://doi.org/10.1007/s10948-015-3302-4}.

\bibitem{NanoCobaltates}
\bibinfo{author}{Guo, H.} \emph{et~al.}
\newblock \bibinfo{title}{Suppression of the outwards-dispersing branches in
  hour-glass magnetic spectra induced by nanoscale phase separation in
  ${\mathrm{la}}_{2\ensuremath{-}x}{\mathrm{sr}}_{x}{\mathrm{coo}}_{4}$}.
\newblock \emph{\bibinfo{journal}{Phys. Rev. B}}
  \textbf{\bibinfo{volume}{100}}, \bibinfo{pages}{014411}
  (\bibinfo{year}{2019}).
\newblock \urlprefix\url{https://link.aps.org/doi/10.1103/PhysRevB.100.014411}.

\bibitem{Ishizaka_2003}
\bibinfo{author}{Ishizaka, K.}, \bibinfo{author}{Taguchi, Y.},
  \bibinfo{author}{Kajimoto, R.}, \bibinfo{author}{Yoshizawa, H.} \&
  \bibinfo{author}{Tokura, Y.}
\newblock \bibinfo{title}{Charge ordering and charge dynamics in
  ${\mathrm{nd}}_{2-x}{\mathrm{sr}}_{x}{\mathrm{nio}}_{4}$
  $(0.33\leq~x\leq~0.7)$}.
\newblock \emph{\bibinfo{journal}{Phys. Rev. B}} \textbf{\bibinfo{volume}{67}},
  \bibinfo{pages}{184418} (\bibinfo{year}{2003}).
\newblock \urlprefix\url{http://link.aps.org/doi/10.1103/PhysRevB.67.184418}.

\bibitem{Uchida_XAS}
\bibinfo{author}{Uchida, M.} \emph{et~al.}
\newblock \bibinfo{title}{Pseudogap-related charge dynamics in the layered
  nickelate ${R}_{2\ensuremath{-}x}$sr${}_{x}$nio${}_{4}$
  ($x\ensuremath{\sim}1$)}.
\newblock \emph{\bibinfo{journal}{Phys. Rev. B}} \textbf{\bibinfo{volume}{86}},
  \bibinfo{pages}{165126} (\bibinfo{year}{2012}).
\newblock \urlprefix\url{http://link.aps.org/doi/10.1103/PhysRevB.86.165126}.

\bibitem{Hu2000}
\bibinfo{author}{Hu, Z.} \emph{et~al.}
\newblock \bibinfo{title}{Hole distribution between the ni $3d$ and o $2p$
  orbitals in
  ${\mathrm{nd}}_{2\ensuremath{-}x}{\mathrm{sr}}_{x}{\mathrm{nio}}_{4\ensuremath{-}\ensuremath{\delta}}$}.
\newblock \emph{\bibinfo{journal}{Phys. Rev. B}} \textbf{\bibinfo{volume}{61}},
  \bibinfo{pages}{3739--3744} (\bibinfo{year}{2000}).
\newblock \urlprefix\url{https://link.aps.org/doi/10.1103/PhysRevB.61.3739}.

\bibitem{Burnus2008a}
\bibinfo{author}{Burnus, T.} \emph{et~al.}
\newblock \emph{\bibinfo{journal}{Phys. Rev. B}} \textbf{\bibinfo{volume}{77}},
  \bibinfo{pages}{125124} (\bibinfo{year}{2008}).

\bibitem{Burnus2008b}
\bibinfo{author}{Burnus, T.} \emph{et~al.}
\newblock \emph{\bibinfo{journal}{Phys. Rev. B}} \textbf{\bibinfo{volume}{77}},
  \bibinfo{pages}{205111} (\bibinfo{year}{2008}).

\bibitem{Mizokawa1995}
\bibinfo{author}{Mizokawa, T.} \& \bibinfo{author}{Fujimori, A.}
\newblock \bibinfo{title}{Unrestricted hartree-fock study of transition-metal
  oxides: Spin and orbital ordering in perovskite-type lattice}.
\newblock \emph{\bibinfo{journal}{Phys. Rev. B}} \textbf{\bibinfo{volume}{51}},
  \bibinfo{pages}{12880--12883} (\bibinfo{year}{1995}).
\newblock \urlprefix\url{https://link.aps.org/doi/10.1103/PhysRevB.51.12880}.

\bibitem{Johnston2014}
\bibinfo{author}{Johnston, S.}, \bibinfo{author}{Mukherjee, A.},
  \bibinfo{author}{Elfimov, I.}, \bibinfo{author}{Berciu, M.} \&
  \bibinfo{author}{Sawatzky, G.~A.}
\newblock \bibinfo{title}{Charge disproportionation without charge transfer in
  the rare-earth-element nickelates as a possible mechanism for the
  metal-insulator transition}.
\newblock \emph{\bibinfo{journal}{Phys. Rev. Lett.}}
  \textbf{\bibinfo{volume}{112}}, \bibinfo{pages}{106404}
  (\bibinfo{year}{2014}).
\newblock
  \urlprefix\url{https://link.aps.org/doi/10.1103/PhysRevLett.112.106404}.

\bibitem{schuesslerLSNO}
\bibinfo{author}{Sch\"{u}{\ss}ler-Langeheine, C.} \emph{et~al.}
\newblock \bibinfo{title}{Spectroscopy of stripe order in
  la$_{1.8}$sr$_{0.2}$nio$_4$ using resonant soft x-ray diffraction}.
\newblock \emph{\bibinfo{journal}{Phys. Rev. Lett.}}
  \textbf{\bibinfo{volume}{95}}, \bibinfo{pages}{156402}
  (\bibinfo{year}{2005}).

\bibitem{Reznik_phonon}
\bibinfo{author}{Reznik, D.} \emph{et~al.}
\newblock \bibinfo{title}{Electron-phonon coupling reflecting dynamic charge
  inhomogeneity in copper oxide superconductors}.
\newblock \emph{\bibinfo{journal}{Nature}} \textbf{\bibinfo{volume}{440}},
  \bibinfo{pages}{1170--1173} (\bibinfo{year}{2006}).
\newblock \urlprefix\url{http://dx.doi.org/10.1038/nature04704}.

\bibitem{Ricci}
\bibinfo{author}{Campi, G.} \emph{et~al.}
\newblock \emph{\bibinfo{journal}{Nature}} \textbf{\bibinfo{volume}{525}},
  \bibinfo{pages}{359} (\bibinfo{year}{2015}).

\bibitem{Fratini}
\bibinfo{author}{Fratini, M.} \emph{et~al.}
\newblock \bibinfo{title}{Scale-free structural organization of oxygen
  interstitials in la$_2$cuo$_{4+y}$}.
\newblock \emph{\bibinfo{journal}{Nature}} \textbf{\bibinfo{volume}{466}},
  \bibinfo{pages}{841–844} (\bibinfo{year}{2010}).
\newblock \urlprefix\url{https://doi.org/10.1038/nature09260}.

\bibitem{Bewley}
\bibinfo{author}{Bewley, R.~I.}, \bibinfo{author}{Guidi, T.} \&
  \bibinfo{author}{Bennington, S.}
\newblock \emph{\bibinfo{journal}{Notiziario Neutroni e Luce di Sincrotrone}}
  \textbf{\bibinfo{volume}{14}}, \bibinfo{pages}{22} (\bibinfo{year}{2009}).

\bibitem{Russina09}
\bibinfo{author}{Russina, M.} \& \bibinfo{author}{Mezei, F.}
\newblock \bibinfo{title}{First implementation of repetition rate
  multiplication in cold, thermal and hot neutron spectroscopy}.
\newblock \emph{\bibinfo{journal}{Nucl. Instrum. Methods in Phys. Rese. Sect.
  A}} \textbf{\bibinfo{volume}{604}}, \bibinfo{pages}{624}
  (\bibinfo{year}{2009}).

\end{thebibliography}

\section*{Acknowledgements (not compulsory)}

The research in Dresden is supported  by  the  Deutsche  Forschungsgemeinschaft through Grant No. 320571839.
D. I. Kh. received support from the Deutsche  Forschungsgemeinschaft via the project SFB 1238 and by the 
German Excellence Initiative. We acknowledge support from the Max Planck-POSTECH-Hsinchu Center for 
Complex Phase Materials.

\end{document}